\documentclass[aps, prmaterials, superscriptaddress, showkeys, twocolumn]{revtex4-2}

\usepackage{graphicx}% Include figure files
\usepackage{dcolumn}% Align table columns on decimal point
\usepackage{bm}% bold math
\usepackage{xspace}
\usepackage{multirow, tabularx}
\usepackage{tabularx}
\usepackage{ulem}

\usepackage[table,xcdraw]{xcolor}
\usepackage{lineno}
\usepackage{hyperref}

\usepackage{hyperref}

\begin{document}

\newcommand{\vts}{V$_x$TaS$_2$\xspace}
\newcommand{\tas}{TaS$_2$\xspace}
\newcommand{\tase}{TaSe$_2$\xspace}
\newcommand{\etal}{\textit{et al.}\xspace}
\DeclareRobustCommand{\emph}[1]{\textit{#1}}

\title{Probing enhanced superconductivity in van der Waals polytypes of V$_x$TaS$_2$}

% \input{authors}
% Authors list for prl style
\author{Wojciech R.~Pudelko}
\email{wojciech.pudelko@psi.ch}
\affiliation{Swiss Light Source, Paul Scherrer Institut, Forschungstrasse 111, Villigen, CH-5232, Switzerland}
\affiliation{Physik-Institut, Universität Zürich, Winterthurerstrasse 190, Zürich, CH-8057, Switzerland}

\author{Huanlong Liu}
\affiliation{Physik-Institut, Universität Zürich, Winterthurerstrasse 190, Zürich, CH-8057, Switzerland}

\author{Francesco Petocchi}
\affiliation{Department of Physics, University of Fribourg, Fribourg, CH-1700, Switzerland}
\affiliation{Department of Quantum Matter Physics, University of Geneva, 24 quai Ernest Ansermet, Geneva, CH-1211, Switzerland}

\author{Hang Li}
\author{Eduardo Bonini Guedes}
\affiliation{Swiss Light Source, Paul Scherrer Institut, Forschungstrasse 111, Villigen, CH-5232, Switzerland}

\author{Julia K\"uspert}
\affiliation{Physik-Institut, Universität Zürich, Winterthurerstrasse 190, Zürich, CH-8057, Switzerland}

\author{Karin von Arx}
\affiliation{Physik-Institut, Universität Zürich, Winterthurerstrasse 190, Zürich, CH-8057, Switzerland}
\affiliation{Department of Physics, Chalmers University of Technology, Göteborg, SE-412 96, Sweden}

\author{Qisi Wang}
\affiliation{Physik-Institut, Universität Zürich, Winterthurerstrasse 190, Zürich, CH-8057, Switzerland}
\affiliation{Department of Physics, The Chinese University of Hong Kong, Shatin, Hong Kong, China}

\author{Ron Cohn Wagner}
\affiliation{Physik-Institut, Universität Zürich, Winterthurerstrasse 190, Zürich, CH-8057, Switzerland}

\author{Craig M.~Polley}
\author{Mats Leandersson}
\author{Jacek Osiecki}
\author{Balasubramanian Thiagarajan}
\affiliation{MAX IV Laboratory, Lund University, 221 00 Lund, Sweden}

\author{Milan Radovi\'c}
\affiliation{Swiss Light Source, Paul Scherrer Institut, Forschungstrasse 111, Villigen, CH-5232, Switzerland}

\author{Philipp Werner}
\affiliation{Department of Physics, University of Fribourg, Fribourg, CH-1700, Switzerland}

\author{Andreas Schilling}
\author{Johan Chang}
\affiliation{Physik-Institut, Universität Zürich, Winterthurerstrasse 190, Zürich, CH-8057, Switzerland}

\author{Nicholas C.~Plumb}
\email{nicholas.plumb@psi.ch}
\affiliation{Swiss Light Source, Paul Scherrer Institut, Forschungstrasse 111, Villigen, CH-5232, Switzerland}

\date{\today}

% \linenumbers

\begin{abstract}
Layered transition metal dichalcogenides (TMDs) stabilize in multiple structural forms with profoundly distinct and exotic electronic phases. Interfacing different layer types is a promising route to manipulate TMDs' properties, not only as a means to engineer quantum devices, but also as a route to explore fundamental physics in complex matter. Here we use angle-resolved photoemission (ARPES) to investigate a strong layering-dependent enhancement of superconductivity in \tas, in which the superconducting transition temperature, $T_c$, of its $2H$ structural phase is nearly tripled when insulating $1T$ layers are inserted into the system. The study is facilitated by a novel vanadium-intercalation approach to synthesizing various \tas polytypes, which improves the quality of ARPES data while leaving key aspects of the electronic structure and properties intact. The spectra show the clear opening of the charge density wave gap in the pure $2H$ phase and its suppression when $1T$ layers are introduced to the system. Moreover, in the mixed-layer $4Hb$ system, we observe a strongly momentum-anisotropic increase in electron-phonon coupling near the Fermi level relative to the $2H$ phase. Both phenomena help to account for the increased $T_c$ in mixed $2H$/$1T$ layer structures.
\end{abstract}


\maketitle

\section{Introduction}

Layered transition metal dichalcogenides (TMDs) are quasi-2D materials with rich phase diagrams involving numerous exotic electronic phases. Tantalum disulfide is a widely studied TMD, whose properties are emblematic of many compounds in this family. It typically stabilizes in one of two structural phases, $2H$ or $1T$. While chemically equivalent, these phases exhibit markedly different electronic properties. The former is a good metal, which enters a charge density wave (CDW) phase at 78 K and becomes a superconductor below 0.8 K \cite{harper1977}. The latter is insulating and undergoes multiple CDW transitions: incommensurate at 543 K, nearly-commensurate at 352 K to fully commensurate (CCDW) below 183 K \cite{rossnagel2011}. In the CCDW phase, the reconstructed unit cell forms a Star-of-David (SoD) cluster, consisting of 13 Ta atoms \cite{wang2020}, which has been proposed to host a quantum spin liquid \cite{klanjsek2017, manas-valero2021}. Despite the 2D character of $1T$-\tas, many theoretical and experimental studies showed non-negligible out-of-plane dispersion \cite{lahoud2014} or stacking dependence \cite{ritschel2015, li2021}, which can further influence its electronic properties and change the gap character from band to Mott insulator \cite{petocchi2022, wang2020, ngankeu2017}.

Due to the delicate balance of interactions, the properties of many TMDs are highly sensitive to minute changes in virtually any material or external parameter, allowing one to, e.g., tune metal-insulator transitions \cite{yu2015}, manipulate magnetic interactions \cite{liu2021}, or induce or enhance superconductivity \cite{liu2021prb, li2017, dong2022, li2012, sipos2008, navaro-mortalla2016}. In addition to these diverse tunable properties, the weak couplings between layers make TMDs an especially attractive platform for engineering exotic electronic phenomena at their interfaces and manufacturing novel quantum devices \cite{yang2018, boix-constant2021, mahajan2022, moriya2020, vano2021}.

One testament to the promise of interfacing various \tas structures is the $4Hb$ phase, a polytype consisting of $1T$ and $1H$/$1H'$ elements (half-layers of the $2H$ structure)  stacked alternately along the $c$-axis \cite{disalvo1973, ribak2020}. This structure has recently attracted attention, due to evidence that it hosts a chiral superconducting state \cite{ribak2020, almoalem2024b, silber2024}. At the same time, the superconducting $T_c$ of the $4Hb$ phase (2.2 K) is nearly triple that of $2H$-\tas. These observations suggest that investigations of mixed 2H/1T-\tas layer systems can yield profound insights into the interplay of superconductivity with topology, dimensionality, many-body interactions, and competing orders.

\begin{figure}[!htp]% vts-structure
    \centering
    \includegraphics[width=8.6cm]{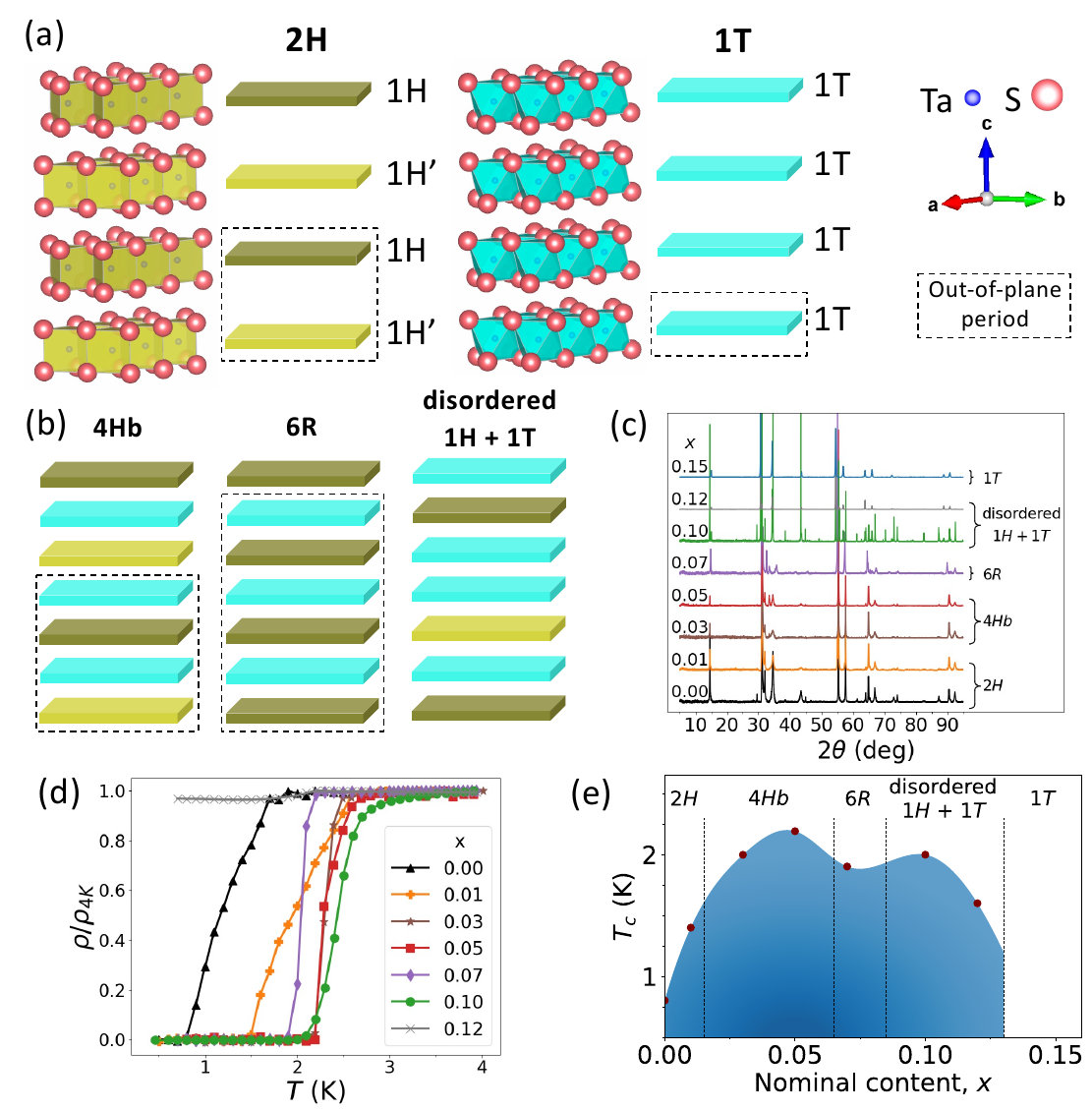}
    \caption{
    (a)~Crystal structures of $2H$- and $1T$-\tas.
    (b)~Illustrations of how the $4Hb$, $6R$, and disordered phases are composed of the $2H$ ($1H/1H'$) and $1T$ layer elements. 
    (c)~Powder XRD measurements of \vts indicating changes in the structure as a function of $x$.
    (d) Normalized resistivity $\rho/\rho_{\mathrm{4K}}$ of \vts for various $x$ values. 
    (e)~$T_c$ as a function of $x$.
    }
    \label{fig:vts-structure}
\end{figure}

This work investigates the strong enhancement of $T_c$ in  $4Hb$ and other mixed-layer phases of \tas. We use angle-resolved photoemission spectroscopy (ARPES) to probe the electronic structure of various polytypes of \tas. Our study employs a novel approach, vanadium intercalation, to obtain the different \tas structural configurations. We find that \vts samples grant clearer ARPES spectra than conventionally-grown \tas samples of the same polytypes, while possessing quantitatively similar band structures \cite{almoalem2024, rossnagel2011} and the same key electronic behaviors, including---most importantly---the same elevated $T_c$ in the $4Hb$ phase relative to $2H$.

The ARPES measurements focus on the $2H$, $4Hb$, and $1T$ phases in order to identify factors at play in the strong enhancement of superconductivity in $4Hb$ and other mixed $2H$/$1T$-\tas systems. Our results show that when $1T$ layers are inserted into the $2H$ structure, as in the $4Hb$ phase, the CDW within the $1H(')$ layers is suppressed while electron-phonon ($e\text{-ph}$) interactions are strongly enhanced. Surprisingly, this enhancement in electron-phonon coupling (EPC) is highly anisotropic in momentum space and only manifested in certain regions of the band structure.\\

\section{Results}

Figure~\ref{fig:vts-structure} presents an overview of the \vts system. The $2H$ and $1T$ layer components and their arrangements in various polytype structures are illustrated in Fig.~\ref{fig:vts-structure}(a). Data from powder x-ray diffraction (XRD) performed on samples with various $x$ values are shown in Fig.~\ref{fig:vts-structure}(b), with labels indicating phases interpreted using standard structural data. Below $x=0.03$, \vts stabilizes in the $2H$ phase. Samples synthesized in the range of about $0.03 \leq x \leq 0.05$ have the $4Hb$ structure. In a narrow intercalation window near $x=0.07$, the structure is assigned to the $6R$ phase. At slightly higher $x$ levels, the arrangement of $1H(')$ and $1T$ layers appears to be disordered. Finally, at $x=0.15$ the compound is in the pure $1T$ phase. Differences in the in-plane lattice constants of V-intercalated and conventional \tas samples are well below 1\%. Larger interlayer spacings (up to 2\%) may be notable, though, particularly as they might promote cleaner sample cleavage and thereby account for the improvements in the quality of ARPES data \cite{supplementary_materials}. The XRD analysis is consistent with core level spectroscopy results, which identify distinct chemical environments of the Ta and S atoms in the $1H(')$ and $1T$ layers \cite{supplementary_materials}.

As shown in Fig.~\ref{fig:vts-structure}(d), a superconducting transition occurs in all samples that contain $1H$ structural components. Intercalating vanadium into the $2H$ ($x=0$) system, and thereby inserting $1T$ layers into the structure, increases the $T_c$. The $T_c$ reaches a maximum of 2.2 K in the $4Hb$ structure ($x=0.05$), which matches the $T_c$ in other $4Hb$-\tas samples \cite{ribak2020}. Further details about the structure and transport properties of the \vts samples are provided in the Supplemental Material (SM) \cite{supplementary_materials}.

\begin{figure}[!htp]% vts-fs
    \centering
    \includegraphics[width=8.6cm]{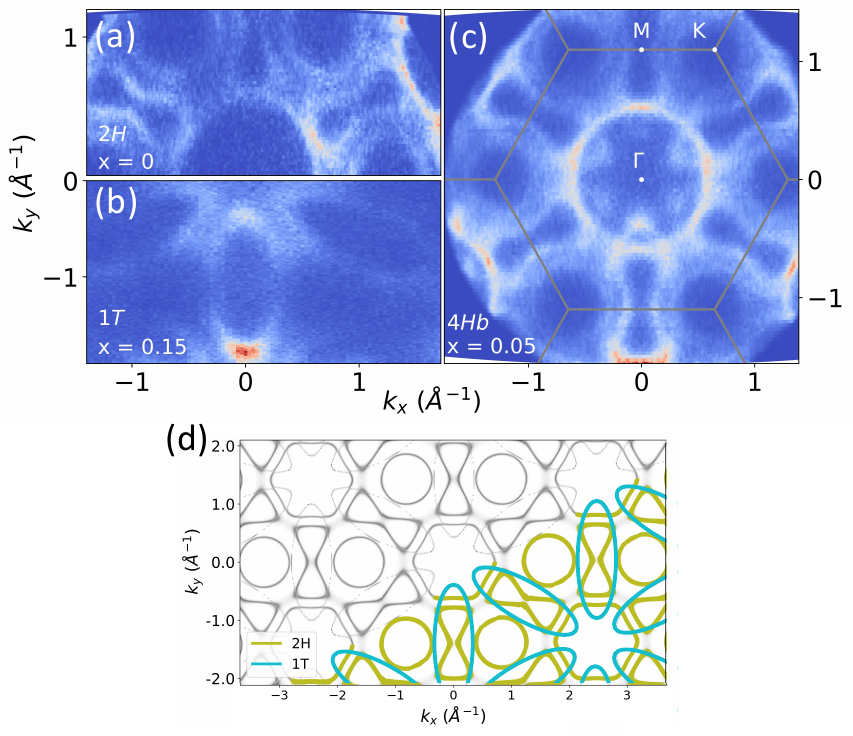}  
    \caption{
    (a)--(c)~ARPES $k$-space maps evaluated at $E-E_F=-100$ meV obtained from $2H$ ($x=0$), $1T$ ($x=0.15$), and $4Hb$ ($x=0.05$) samples, respectively.
    (d)~DMFT calculation of a $2H+1T$ model system, evaluated at $E-E_F=-100$~meV. The overlaid contours indicate contributions of the $2H$ and $1T$ components, which are evident in (c).
    }
    \label{fig:vts-fs}
\end{figure}

Figures~\ref{fig:vts-fs}(a)--(c) show ARPES momentum maps in the surface $k_x\text{-}k_y$ plane, evaluated 100 meV below $E_F$ for the $x=0$, 0.05, and 0.15 samples. The electronic structures of the $x=0$ and 0.15 samples are consistent with results reported for the $2H$ and $1T$ structures, respectively \cite{rossnagel2011, wijayaratne2017, wang2020, clerc2004}. The $2H$ phase [Fig.~\ref{fig:vts-fs}(a)] is a metallic system consisting of two so-called ``barrel''-shaped Fermi surface (FS) sheets centered around the $\Gamma$ and $K$ points, plus a ``dogbone'' sheet located at the $M$ point \cite{wijayaratne2017, blaha1991}, labeled as $2H\text{-B}$ and $2H\text{-D}$, respectively. The $1T$ phase [Fig.~\ref{fig:vts-fs}(b)] comprises a single gapped band that, when viewed at the energy of the gap, forms a ``flower petal'' in the $k_x\text{-}k_y$ plane centered around $M$ ($1T\text{-FP}$). In the $x=0.05$ sample [Fig.~\ref{fig:vts-fs}(c)], spectral features originating from both the $2H$ and $1T$ phases are visible, consistent with data reported on $4Hb$-\tas \cite{ribak2020}. The contributions of the $2H$ and $1T$ components to the band structure of the mixed-layer phases are highlighted by density mean field theory (DMFT) calculations, which consider a $1T$ layer inserted into the $2H$ structure. Unlike density functional theory, the DMFT method accounts for a Hubbard interaction energy, $U$, within the $1T$ layers. The model results, shown in Fig.~\ref{fig:vts-fs}(d), are in reasonable agreement with the ARPES data from the $4Hb$ ($x=0.05$) phase when assuming $U=0.4$ eV, similar to the $1T$ system \cite{petocchi2022}.

\begin{figure}[!htbp]% vts-bm
    \centering
    \includegraphics[width=8.6cm]{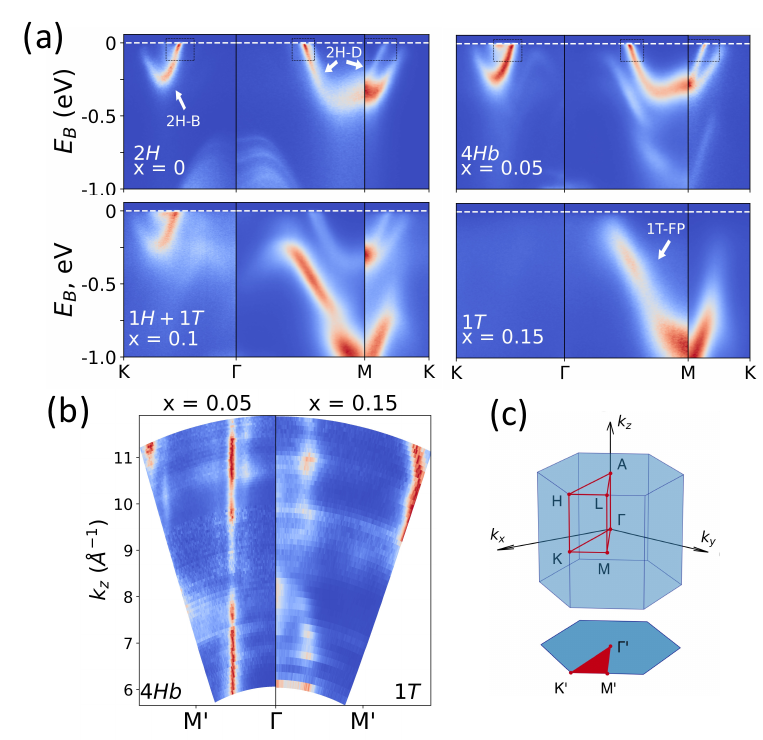}  
    \caption{
    (a)~ARPES measurements of electron dispersions in $2H$, $4Hb$, mixed $1H+1T$, and $1T$ samples with the noted $x$ values. Barrel, dogbone, and flower-petal bands are labeled as $2H\text{-B}$, $2H\text{-D}$ and $1T\text{-FP}$, respectively. Rectangles in the spectra of the $2H$ and $4Hb$ samples highlight regions that are analyzed in detail in Fig.~\ref{fig:vts-kinks}.
    (b)~ARPES $k_x\text{-}k_z$ maps of $4Hb$ and $1T$ samples, evaluated 100~meV below $E_F$.
    (c)~Bulk and surface Brillouin zones. For simplicity, the ``prime'' notation of the surface Brillouin zone labels is dropped when considering in-plane results.
    }
    \label{fig:vts-bm}
\end{figure}

Figure~\ref{fig:vts-bm}(a) shows the evolution of the band dispersions in \vts with increasing $x$. The top left panel displays the $2H$ ($x=0$) phase, consisting of $2H\text{-B}$ and $2H\text{-D}$ bands. The $1T$ ($x=0.15$) phase, consisting of the $1T\text{-FP}$ band, is shown in the bottom right panel. In the mixed $4Hb$ ($x=0.05$) and $1H+1T$ ($x=0.1$) systems, all three bands are visible. While the most salient features of the $2H$- and $1T$-derived bands are preserved in these phases, there are notable differences among the spectra, which could be points of future investigation. These include a large downward shift in the energy of the $2H$ valence bands around $\Gamma$, and faint replicas of the $1T\text{-FP}$ band in the $4Hb$ phase. In addition, compared to the pure $2H$ ($x=0$) phase, the $4Hb$ ($x=0.05$) and disordered $1H+1T$ ($x=0.1$) structures show a slight upward shift of the $2H\text{-B}$ and -D conduction bands. The corresponding changes in Fermi momenta appear consistent with a recent observation of charge transfer between the $1T$ and $1H(')$ layers within $4Hb$-\tas \cite{almoalem2024}.

We also note that, although we observe signatures of SoD reconstruction in $1T$-\vts ($x=0.15$), its ARPES spectrum does not exhibit a shallow, gapped spectral feature at $\Gamma$, which is seen in conventionally-grown $1T$ samples and widely ascribed to a Mott gap \cite{lahoud2014, ritschel2015, wang2021}. The lack of this feature is attributable to a difference in the vertical stacking alignment of layers \cite{ritschel2015, wang2020, petocchi2022}. As the ``Mott gap'' feature does not carry over into the $4Hb$ system where we investigate the enhancement of $T_c$ \cite{almoalem2024, ribak2020}, its absence in $1T$-\vts does not affect the current study.

Varying the photon energy in ARPES allows the electronic structure to be probed as a function of the out-of-plane momentum, $k_z$. Figure~\ref{fig:vts-bm}(b) compares momentum maps from the $x=0.05$ ($4Hb$, left) and 0.15 ($1T$, right) samples in the $k_x\text{-}k_z$ plane, evaluated 100~meV below $E_F$. Here the $1T\text{-FP}$ band in the $x=0.15$ sample exhibits weak but observable momentum dependence along $k_z$, similar to previous reports \cite{ngankeu2017, jung2022, disante2017, vydrova2015, sato2004, weber2018, strocov2012}. By contrast, the same $1T$-derived bands in the $4Hb$ ($x=0.05$) phase show little to no variation with respect to $k_z$, signaling enhanced two-dimensionality. For orientation, the bulk and surface Brillouin zones are sketched in Fig.~\ref{fig:vts-bm}(c).

\begin{figure*}[!htbp]% vts-kinks
    \centering
    \includegraphics[width=17.2cm]{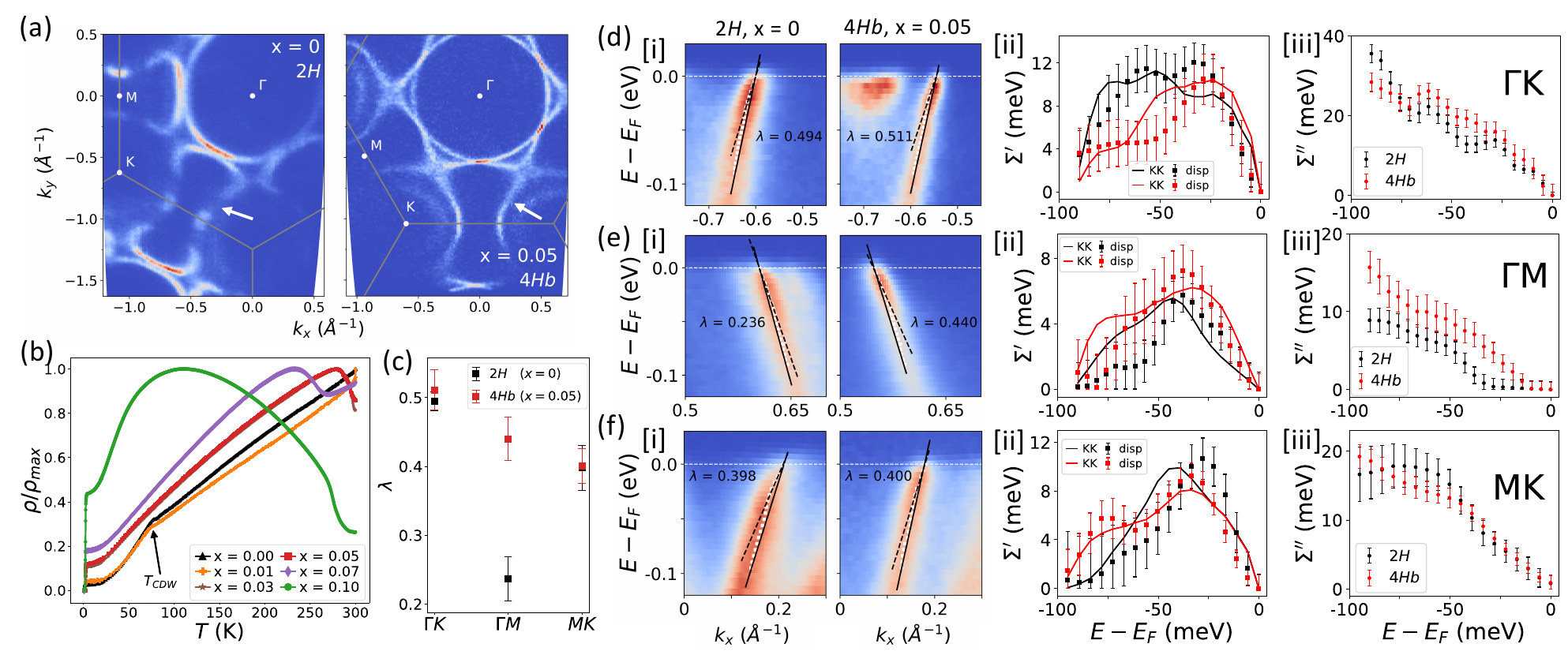}  
    \caption{
    (a)~FSs of $2H$ ($x=0$, left panel) and $4Hb$ ($x=0.05$, right panel) samples measured at 18 K. Gapped portions of the $2H$ FS (arrows) are signatures of CDW order. 
    (b)~Normalized resistivity measurements as a function of $x$. A cusp in the resistivity due to the CDW transition at $T\sim76$ K in the $2H$ phase ($x = 0$ and 0.01) is suppressed in the $4Hb$ phase and other mixed polytypes ($x\geq0.05$). 
    (c)~EPC constant $\lambda$ of the $2H$ ($x=0$) and $4Hb$ ($x=0.05$) samples along the $\Gamma\text{--}K$, $\Gamma\text{--}M$ and $M\text{--}K$ directions. 
    (d)~[i] Fitted band dispersion (white squares), as well as the renormalized Fermi velocity (dashed black line) and extracted bare band velocity (solid black line) along the $\Gamma\text{--}K$ direction for the $x = 0$ (right panel) and 0.5 samples (left panel).
    [ii] ([iii]) Real (imaginary) part of the self-energy, $\Sigma'$ ($\Sigma''$), along $\Gamma\text{--}K$ for $x=0$ and 0.05. Two methods of computing $\Sigma'(\omega)$ are shown: Kramers-Kronig transformation of $\Sigma''(\omega)$  (KK) and subtraction from the bare band dispersion (disp).
    Panels (e) and (f) show the same analysis as in (d), applied along the $\Gamma\text{--}M$ and $M\text{--}K$ directions, respectively.
    }
    \label{fig:vts-kinks}
\end{figure*}

The FS presented in Fig.~\ref{fig:vts-kinks}(a) (left panel) exhibits well-defined gapped regions along the dogbone sheets---clear signatures of a CDW reconstruction. The locations of the gapped portions in the FS (indicated with the white arrows) match those of the sister compound $2H$-\tase, where the $q_{\mathrm{CDW}}=2/3|\Gamma\mathrm{-}M|$ ordering wavevector leads to a commensurate $3\times 3$ reconstruction of the unit cell \cite{borisenko2008, inosov2008, li2018}. Analysis of the Lindhard susceptibility of $2H$-\tas, included in the SM \cite{supplementary_materials}, further establishes the similarities between its CDW state and that of $2H$-\tase. Moving from the $2H$ ($x=0$) system to the $4Hb$ structure ($x=0.05$), the CDW instability is suppressed, as seen in the right panel of Fig.~\ref{fig:vts-kinks}(a). A comparison of the energy distribution curves (EDCs) taken at the $k$-points where the CDW gap opens is included in the SM \cite{supplementary_materials}. The suppression of the CDW phase transition in samples containing a mixture of $2H$ and $1T$ layers ($0.03 \leq x \leq 0.10$) is confirmed by resistivity measurements, as shown in Fig.~\ref{fig:vts-kinks}(b). 

In addition, the ARPES data show signatures of electron interactions in the $2H$ band structure in the form of a dispersion ``kink'' anomaly roughly 30 meV below $E_F$. The kink can be seen in ARPES dispersion cuts along various momentum directions. It is highlighted by the boxes in Fig.~\ref{fig:vts-bm}(a) and examined in detail in Figs.~\ref{fig:vts-kinks}(d)--(f)[i]. This energetically sharp feature, which has also been noted in previous work \cite{wijayaratne2017}, ostensibly signals that the electrons couple to a particular boson mode at the same energy. We ascribe the kink to $e\text{-ph}$ interactions; calculations of the $2H$-\tas phonon band structure find an optical phonon branch with a flat dispersion at the same $\sim30$ meV scale, where there is substantial EPC over an extended region centered around the $\Gamma\text{--}M$ wavevector \cite{hinsche2018}.

In Figs.~\ref{fig:vts-kinks}(d)--(f), we analyze the dispersion kink to assess changes in the EPC parameter, $\lambda$, between the $2H$ ($x=0$) and $4Hb$ ($x=0.05$) samples. We evaluate $\lambda$ from analysis of the complex electronic self-energy, $\Sigma(\omega)=\Sigma'(\omega)+i\Sigma''(\omega)$, where $\omega=E-E_F$. We follow the general approach of Refs.~\cite{kordyuk2005, wen2018}, which is based on Lorentzian fitting of the ARPES momentum distribution curves (MDCs) a fixed energies. For each sample, $\Sigma(\omega)$ is analyzed along the $\Gamma\text{--}K$, $\Gamma\text{--}M$ and $M\text{--}K$ directions. $\Sigma'(\omega)$ and $\Sigma''(\omega)$ are related to the spectra as the difference between the renormalized and non-interacting ``bare'' dispersions, and from the widths of the MDCs at the corresponding energies, respectively. The methodology involves extracting $\Sigma'(\omega)$ by two independent methods---Kramers-Kronig (KK) transformation of $\Sigma''(\omega)$ and analysis of the dispersion (disp)---which self-consistently determine the non-interacting Fermi velocity, $v_F^0$, and thus the absolute scaling of $\Sigma(\omega)$. Details of the calculations and results are described in the SM \cite{supplementary_materials}.

The extracted $\lambda$ values are presented in Fig.~\ref{fig:vts-kinks}(c). For the $2H$ phase ($x=0$), we find $\lambda$ ranges from 0.236 along the $\Gamma-M$ direction to 0.494 along $\Gamma-K$. This variation in $\lambda$ is not surprising, as the EPC is known to be $k$-dependent in $2H$-\tas \cite{wijayaratne2017, lian2022}. It is remarkable, however, that the \textit{enhancement} in EPC in going from the $2H$ to the $4Hb$ phase ($x=0.05$) is strongly momentum-dependent. Namely, $\lambda$ is virtually unchanged at the measured points along the $\Gamma\text{--}K$ and $M\text{--}K$, while a large enhancement in the interactions is found along the $\Gamma\text{--}M$ direction, where $\lambda=0.440$ (an increase of about 85\%).\\

\section{Discussion}

The results here offer spectroscopic insights into key factors that lead to a strong enhancement of $T_c$ in the mixed-layer polytypes. The suppression of CDW order restores states to the FS that are especially likely to couple strongly to phonons, allowing them to form Cooper pairs in the competing superconducting phase. Meanwhile the enhancement in EPC at the $\sim30$ meV scale, while strongly momentum-anisotropic, is nonetheless substantial in terms of its likely impact on $T_c$. As an illustration, by applying the McMillan equation \cite{mcmillan1968} and averaging the $\lambda$ values extracted along the three momentum directions in Figs.~\ref{fig:vts-kinks}(d)--(f), the measured $T_c$'s of 0.8 K and 2.2 K in the $2H$ and $4Hb$ systems, respectively, can be rationalized under reasonable assumptions, even before other factors are taken into account \cite{supplementary_materials}.

It follows from this that Mott-like interactions originating from the $1T$ planes should not play a leading role in elevating the $T_c$ in $4Hb$-\tas. This aligns with the negligible $k_z$ dispersion of the $1T\text{-FP}$ band within the $4Hb$ system [Fig.~\ref{fig:vts-bm}(b)], as well as the similar $U$ values inferred from DMFT modeling of $1T$ layers in pure and $4Hb$-like systems, which both suggest that Mott interactions remain largely confined within the $1T$ planes. The ARPES measurements furthermore suggest that changes in carrier density, either due to inter- or intralayer charge transfer or doping by the intercalants, probably have little \textit{direct} impact on $T_c$. Observed changes in the FS volume of the $2H$ bands are too small to account for the large $T_c$ enhancement via the density of states alone and, if anything, would tend to reduce the states at $E_F$ and hence the $T_c$ \cite{almoalem2024}. Still, small charge transfer and/or doping effects could indirectly influence superconductivity in profound ways, e.g., by altering the $e\text{-ph}$ interactions. 

The strongly momentum-dependent increase in $\lambda$ highlights the complexity of the interactions in these systems and the need for further study. It is tempting to draw inferences from the fact that we observe a dramatic EPC enhancement at a point in the dispersion along the same momentum direction ($\Gamma\text{--}M$) as the $2H$ CCDW ordering wavevector. It is possible the the same phonon interactions at play in the CDW order of the $2H$ phase are strengthened and bolster superconductivity in the $4Hb$ system, similar to recent findings in charge-ordered cuprates \cite{wang2021}. Drawing such conclusions, however, will require further studies to the identify the phonon wavevectors where EPC is strongly enhanced. 

The observations here may have broader relevance outside of the \tas bulk polytypes. Suppression of CDW order and enhancement of EPC were also observed in other intercalated TMDs \cite{hu2007} and atomically thin layers of $2H$-\tas \cite{navaro-mortalla2016, yang2018}. The similar behaviors in these systems may be general consequences of isolating individual $1H(')$ layers, which may reduce metallic screening and alter the phonon dispersions and FS nesting conditions.\\

\section{Conclusions}

In conclusion, using ARPES, we have probed the electronic structures of $2H$-, $1T$-, and $4Hb$-\tas in order to investigate the strong enhancement of superconductivity in mixed $2H$/$1T$ layer polytypes relative to the pure $2H$ structure. Our work utilized vanadium intercalation as a novel route to synthesize high-quality \tas polytypes. The ARPES measurements demonstrate the clear presence of the CDW gap in pure $2H$-\tas, which by a comparison to isovalent $2H$-\tase, might be associated with a commensurate 3$\times$3 reconstruction. ARPES and transport data further show that the CDW is suppressed in mixed $2H$/$1T$ layer forms of \vts ($0.03 \leq x \leq 0.1$). The spectra exhibit a kink dispersion anomaly in the $2H$-derived bands as a signature of energetically sharp electron excitations---presumably $e\text{-ph}$ interactions---roughly 30 meV below $E_F$. In the $4Hb$ structure ($x=0.05$), the EPC associated with the kink is strongly enhanced in a highly momentum-dependent manner. Both the suppression of the CDW and the momentum-anisotropic increase of EPC are likely to be key factors in explaining the large enhancement of $T_c$ when $2H$-\tas layers or half-layers are electronically isolated---either as polytypes incorporating $1T$ layers ($4Hb$, $6R$, and other phases), or in monolayer or true heterostructure form. \\ 

\section*{Acknowledgments}

W.R.P., N.C.P., J.K., K.v.A., and J.C.~acknowledge support from the Swiss National Science Foundation through Project Numbers 200021\_185037 and 200021\_188564. Q.W.~acknowledges support by the Research Grants Council of Hong Kong (ECS No.~24306223). We acknowledge MAX IV Laboratory for time on Beamline Bloch under Proposal 20230375. Research conducted at MAX IV, a Swedish national user facility, is supported by the Swedish Research council under contract 2018-07152, the Swedish Governmental Agency for Innovation Systems under contract 2018-04969, and Formas under contract 2019-02496. The authors thank A.~Kanigel for helpful discussions.\\

\bibliography{bib}

\end{document}

\newcommand{\vts}{${\mathrm{V}_{x}\mathrm{TaS}_{2}}$\xspace}
\newcommand{\tas}{TaS$_2$\xspace}
\newcommand{\tase}{TaSe$_2$\xspace}
\DeclareRobustCommand{\emph}[1]{\textit{#1}}

\title{Supplemental Material \\ 
Probing enhanced superconductivity in van der Waals polytypes of V$_x$TaS$_2$}

% Authors list for prl style
\author{Wojciech R.~Pudelko}
\email{wojciech.pudelko@psi.ch}
\affiliation{Swiss Light Source, Paul Scherrer Institut, Forschungstrasse 111, Villigen, CH-5232, Switzerland}
\affiliation{Physik-Institut, Universität Zürich, Winterthurerstrasse 190, Zürich, CH-8057, Switzerland}

\author{Huanlong Liu}
\affiliation{Physik-Institut, Universität Zürich, Winterthurerstrasse 190, Zürich, CH-8057, Switzerland}

\author{Francesco Petocchi}
\affiliation{Department of Physics, University of Fribourg, Fribourg, CH-1700, Switzerland}
\affiliation{Department of Quantum Matter Physics, University of Geneva, 24 quai Ernest Ansermet, Geneva, CH-1211, Switzerland}

\author{Hang Li}
\author{Eduardo Bonini Guedes}
\affiliation{Swiss Light Source, Paul Scherrer Institut, Forschungstrasse 111, Villigen, CH-5232, Switzerland}

\author{Julia K\"uspert}
\affiliation{Physik-Institut, Universität Zürich, Winterthurerstrasse 190, Zürich, CH-8057, Switzerland}

\author{Karin von Arx}
\affiliation{Physik-Institut, Universität Zürich, Winterthurerstrasse 190, Zürich, CH-8057, Switzerland}
\affiliation{Department of Physics, Chalmers University of Technology, Göteborg, SE-412 96, Sweden}

\author{Qisi Wang}
\affiliation{Physik-Institut, Universität Zürich, Winterthurerstrasse 190, Zürich, CH-8057, Switzerland}
\affiliation{Department of Physics, The Chinese University of Hong Kong, Shatin, Hong Kong, China}

\author{Ron Cohn Wagner}
\affiliation{Physik-Institut, Universität Zürich, Winterthurerstrasse 190, Zürich, CH-8057, Switzerland}

\author{Craig M.~Polley}
\author{Mats Leandersson}
\author{Jacek Osiecki}
\author{Balasubramanian Thiagarajan}
\affiliation{MAX IV Laboratory, Lund University, 221 00 Lund, Sweden}

\author{Milan Radovi\'c}
\affiliation{Swiss Light Source, Paul Scherrer Institut, Forschungstrasse 111, Villigen, CH-5232, Switzerland}

\author{Philipp Werner}
\affiliation{Department of Physics, University of Fribourg, Fribourg, CH-1700, Switzerland}

\author{Andreas Schilling}
\author{Johan Chang}
\affiliation{Physik-Institut, Universität Zürich, Winterthurerstrasse 190, Zürich, CH-8057, Switzerland}

\author{Nicholas C.~Plumb}
\email{nicholas.plumb@psi.ch}
\affiliation{Swiss Light Source, Paul Scherrer Institut, Forschungstrasse 111, Villigen, CH-5232, Switzerland}

\date{\today}

\maketitle

\section{Crystal growth and characterization} 
High-quality crystalline \vts samples were synthesized by the chemical vapor transport method with iodine as a transport agent. Stoichiometric amounts of the raw materials (99.9\% V, 99.99\% Ta, and 99.9\% S powders) and iodine (density: 2.0 mg/cm$^3$) were mixed and heated at 800 \textdegree{}C for three weeks in evacuated quartz tubes. Then they were slowly cooled down to 600 \textdegree{}C with a cooling rate of 2~\textdegree{}C/h. Finally, the tubes were cooled down in the furnace by auto-switching off the power. 

\begin{figure*}[!hb]% vts-lattice_consts-supplement
    \centering
    \includegraphics[width=0.75\linewidth]{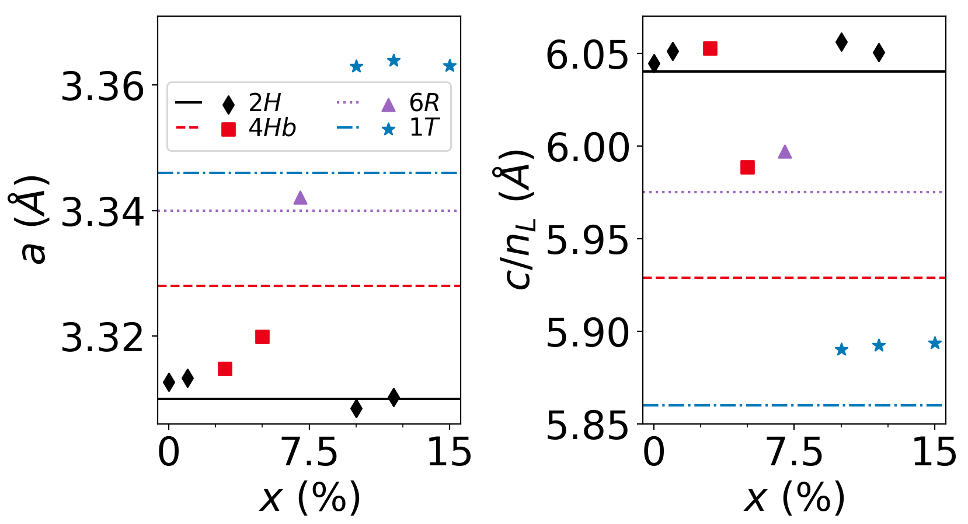}
    \caption{Changes in unit cell parameters with increasing vanadium content $x$ (in \%). Values of the out-of-plane lattice constant $c$ have been divided by the number of \tas layers within a single unit cell---i.e., $n_L = 2$, 4, 6, 1 for $2H$, $4Hb$, $6R$ and $1T$ phases, respectively. For samples in the disordered $1H+1T$ range ($x=10\%$ and 12\%), $c/n_L$ values are plotted for both $n_L=2$ and $n_L=1$. Lines indicate lattice parameters for “standard” (i.e., non-intercalated) \tas samples in the same phases, taken from Refs. \cite{meyer1975, pal2023}.}
    \label{fig:vts-sm-lattice_consts}
\end{figure*}

The crystal structures were characterized by powder x-ray diffraction, collected on all the as-prepared crystalline \vts samples using a Stoe STADIP diffractometer at room temperature with Cu K$\alpha_1$ radiation ($\lambda =1.5418$~\AA). Diffraction results are summarized in Fig.~\ref{fig:vts-sm-lattice_consts} and Table~\ref{tab:lattice_consts}.

\begin{table}[!ht]\caption{Values of the lattice constants plotted in Fig.~\ref{fig:vts-sm-lattice_consts} and relative changes with respect to the same conventionally-grown (i.e., non-intercalated) phases. That is to say, $\Delta x_i = (x_i - x_0) / x_0 \cdot 100\%$, where $x_i$ is a lattice parameter ($a$ or $c$) of the V-intercalated sample, and $x_0$ is corresponding value for the pure \tas compound of the same structural phase.}
\label{tab:lattice_consts}
\begin{ruledtabular}
\begin{tabularx}{\linewidth}{cc|ccc|ccc}
x   & phase     & $a_0$, \AA     & $a$, \AA    & $\Delta a$, \%    & $c_0$, \AA     & $c$, \AA    & $\Delta c$, \% \\ \hline
0                     & \multirow{2}{*}{2H}  & \multirow{2}{*}{3.310} & 3.3127 & 0.082  & \multirow{2}{*}{12.08} & 12.089 & 0.076  \\
0.01                  &                      &                        & 3.3133 & 0.100  &                        & 12.102 & 0.185  \\ \hline
0.03                  & \multirow{2}{*}{4Hb} & \multirow{2}{*}{3.328} & 3.3148 & -0.397 & \multirow{2}{*}{23.72} & 24.211 & 2.086  \\
0.05                  &                      &                        & 3.3199 & -0.243 &                        & 23.954 & 1.004  \\ \hline
0.07                  & 6R                   & 3.340                  & 3.3421 & 0.063  & 35.85                  & 35.983 & 0.372  \\ \hline
\multirow{2}{*}{0.1}  & 1H($'$)              & 3.310                  & 3.3085 & -0.045 & 6.04                   & 6.056  & 0.266  \\
                      & 1T                   & 3.346                  & 3.3630 & 0.508  & 5.86                   & 5.890  & 0.515  \\
\multirow{2}{*}{0.12} & 1H($'$)              & 3.310                  & 3.3103 & 0.009  & 6.04                   & 6.051  & 0.174  \\
                      & 1T                   & 3.346                  & 3.3639 & 0.535  & 5.86                   & 5.893  & 0.555  \\ \hline
0.15                  & 1T                   & 3.346                  & 3.3631 & 0.511  & 5.86                   & 5.894  & 0.575 
\end{tabularx}
\end{ruledtabular}
\end{table}

\begin{figure*}[!htb]% vts-hysteresis
    \centering
    \includegraphics[width=0.45\linewidth]{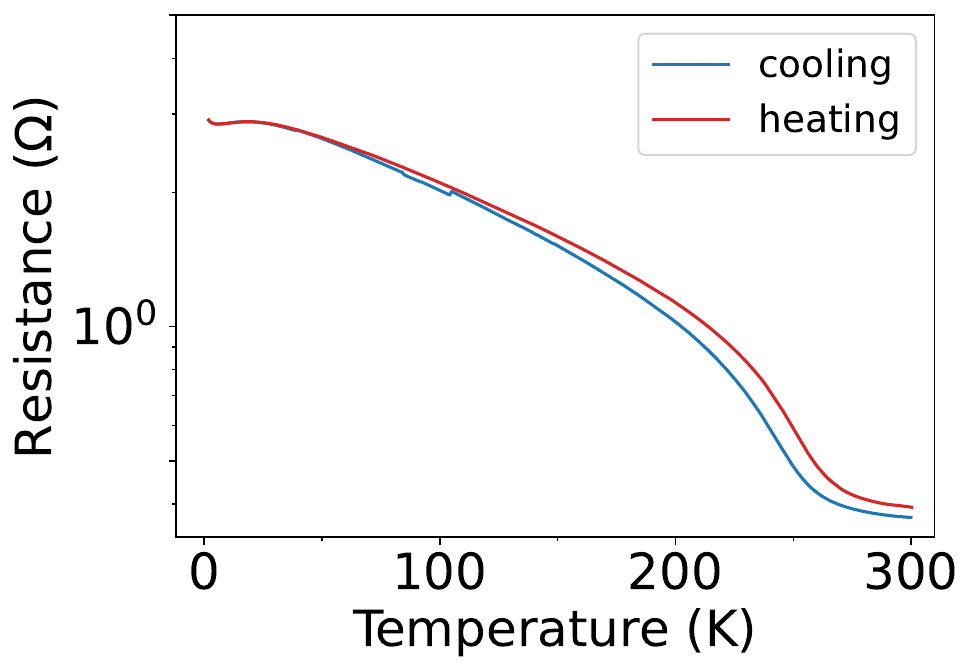}
    \caption{Resistance of a $1T$-\vts sample ($x=0.15$) measured as a function of temperature. Both cooling and heating cycles are shown.}
    \label{fig:vts-sm-hysteresis}
\end{figure*}

The electrical transport was measured using a Quantum Design physical property measurement system equipped with a {}$^{3}$He insert and a standard four-point probe. The samples were cut into rectangular shapes and four electrical contacts were applied to each sample with 25-$\mu$m-diameter Au wires attached to the samples using Dupont 4929 silver paint. In addition to transport measurements shown in the main text, in Fig.~\ref{fig:vts-sm-hysteresis} we plot the resistance of $1T$-\vts ($x=0.15$) as a function of temperature. The temperature dependence of the resistivity in $1T$-\vts samples is qualitatively similar to conventionally-grown $1T$-\tas insofar as both are insulating and show a hysteresis loop in the transition between nearly-commensurate and commensurate CDW phases \cite{wang2020}. However, in $1T$-\vts, the transition is centered close to 250 K, about 50 K higher than in conventional $1T$-\tas. The hysteresis loop is also narrower in $1T$-\vts (about 10 K compared to 30 K). The overall shapes of the insulating resistance curves also differ. The differences in the transport behaviors of vanadium-intercalated and conventionally-grown $1T$-\tas samples have no impact on the present study's conclusions. We speculate that the discrepancies may relate to the larger interlayer spacing and/or possibly modified layer stacking in $1T$-\vts samples compared to conventional $1T$-\tas.

\begin{figure*}[!htb]% vts-xps
    \centering
    \includegraphics[width=0.55\linewidth]{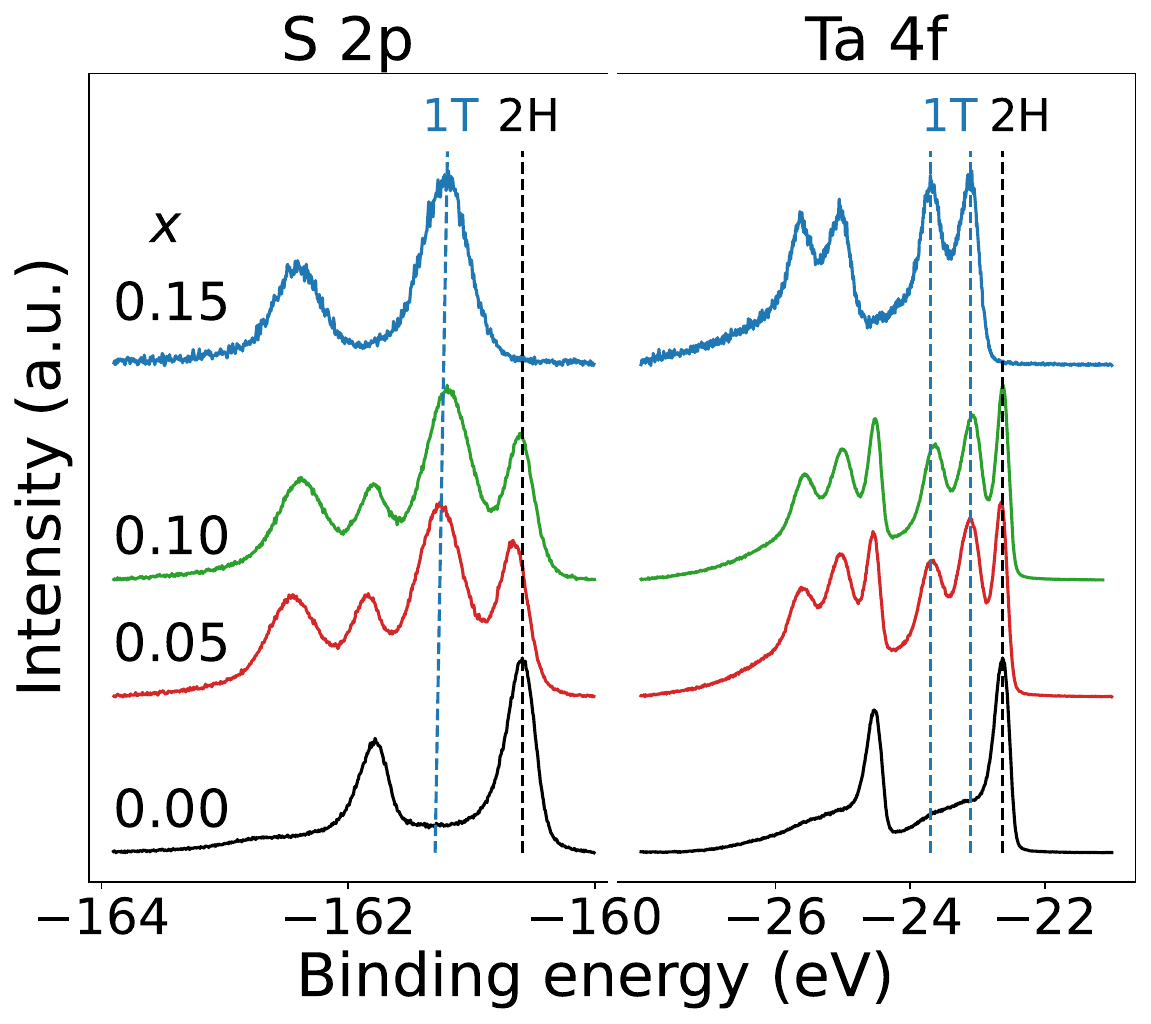}
    \caption{Core level XPS spectra showing the coexistence of local $2H$- and $1T$-like atomic environments for samples with mixed-layer structures.}
    \label{fig:vts-sm-xps}
\end{figure*}

The chemical environments of the Ta and S elements were examined using x-ray photoelectron spectroscopy (XPS). Results from core level spectra are consistent with the XRD analysis. Figure~\ref{fig:vts-sm-xps} presents S $2p$ and Ta $4f$ core level spectra as a function of $x$. The spectra were acquired with 417 eV and 273~eV photon energies, respectively, so that their photoelectron kinetic energies (and hence probing depths) are matched. In pure $2H$ ($x=0$) samples, both core level spectra appear as pure doublets. In $1T$ samples ($x=0.15$), the Ta $4f$ spectrum is composed of two doublets, which are understood to arise due to distinct Ta environments in the Star-of-David ordered phase \cite{ribak2020}. Meanwhile, $x=0.05$ and 0.1 samples exhibit a superposition of the $2H$ and $1T$ spectra, consistent with the presence of both layer elements \cite{hughes1976, scarfe1989} and similar to previous XPS measurements of the $4Hb$ structure \cite{ribak2020}.

\section{Angle-Resolved Photoemission Spectroscopy} 
Angle-resolved photoemission spectroscopy (ARPES) experiments were performed at the SIS and Bloch beamlines of the Swiss Light Source and MAX-IV synchrotrons, respectively. Data were obtained with horizontally polarized light at photon energies of 75 and 130 eV. Scans along the surface-perpendicular momentum direction, $k_z$, were measured with photon energies ranging between 30 and 130 eV. The Fermi level was measured from polycrystalline gold cleaned by ion sputtering. Single crystals were cleaved in ultra high vacuum and studied under pressure better than 10$^{-10}$~mbar. Measurements were performed on samples with vanadium content $x = 0$, 0.05, 0.1, and 0.15, at temperatures between 10--18~K.

\begin{figure*}[!htb]% vts-termination
    \centering
    \includegraphics[width=0.55\linewidth]{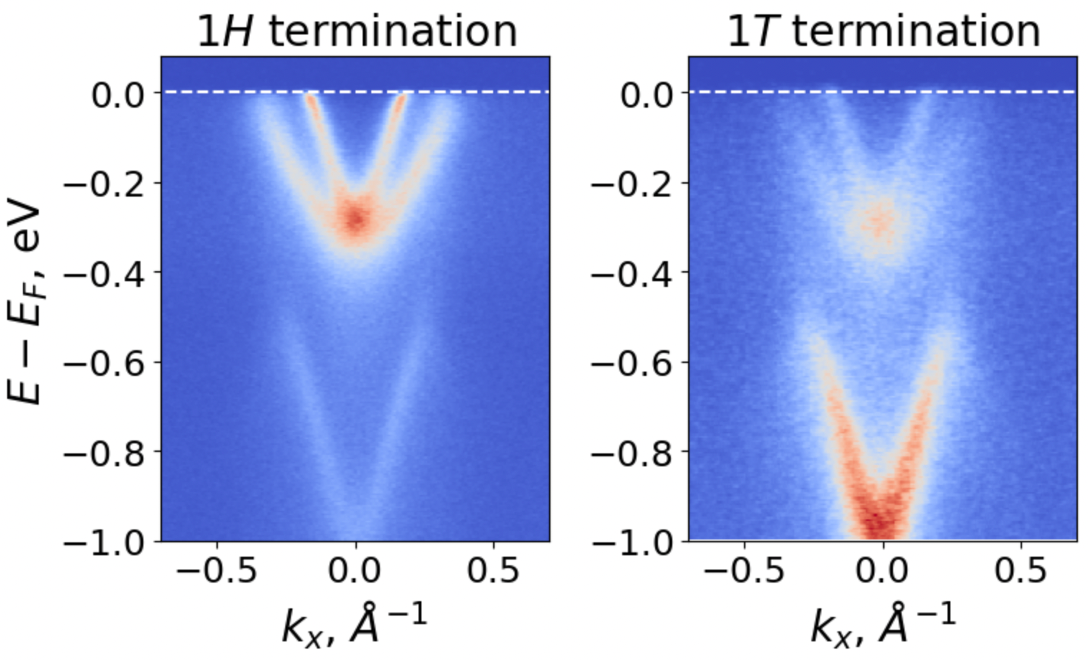}
    \caption{ARPES spectra illustrating variations in spectral weight between bands originating from different ($1H$/$1T$) structural components, suggesting termination effects in mixed polytypes. The band maps correspond to the $x=0.05$ sample, acquired along the $M$-$K$ direction.}
    \label{fig:vts-sm-termination}
\end{figure*}

In the case of the mixed polytypes, the surface termination can differ based on the crystal region and potentially have an influence on the obtained results. Although an unambiguous identification of the type of top-most layer would require a real-space probe, some information can be extracted from the ARPES spectra. Figure~\ref{fig:vts-sm-termination} shows a comparison between two band maps acquired from the $x=0.05$ samples. The data exhibit clear dissimilarities in the spectral weight between bands originating from different layer components, suggesting different surface terminations. Based on this observation, the data analyzed in the main text are inferred to originate from spots corresponding to the $1H$-termination.

\section{DMFT calculations} 
In order to construct a tight-binding (TB) model of $2H$-\tas, we parameterize the two density functional theory (DFT) bands crossing the Fermi level reported in Ref.~\cite{li2017} as

\begin{equation}
\varepsilon_{\alpha}\left(\mathbf{k}\right)=\varepsilon_{\alpha}^{0}+\sum_{\mathbf{R}}t_{\alpha}\left(\mathbf{R}\right)\cos\left(\mathbf{k}\cdot\mathbf{R}\right),
\end{equation}

\noindent
where $\alpha=\left\{ 1,2\right\}$  is the band index, $\varepsilon_{\alpha}^{0}$ an energy level, $\mathbf{R}$ denotes the vectors of the 2D hexagonal lattice, and $t_{\alpha}\left(\mathbf{R}\right)$ the TB parameters which we retained up to the twentieth neighbor. This setup, employed also in Ref.~\cite{laverock2013}, describes the two molecular orbitals of a $2H$-\tas bilayer that cannot be associated to a single Ta ion. We therefore rotate $\varepsilon_{\alpha}\left(\mathbf{k}\right)$ to the basis where the local energy levels are degenerate. This yields a different Hamiltonian representation with the same eigenvalues and dispersion $\epsilon_{ab}\left(\mathbf{k}\right)$,  where $a$ is the index identifying the Ta ion. We performed the same fitting procedure on the DFT band structure of $1T$-\tas in the undistorted phase reported in Ref.~\cite{shao2016prb}. In the latter case, only one band is present and a change of representation is not needed. The flexibility of this approach allowed us to build the TB model associated with different $2H$ stacking arrangements perturbed by $1T$ intercalation.\\

\begin{figure*}[!htbp]% vts-cdw-supplement
    \centering
    \includegraphics[width=0.95\linewidth]{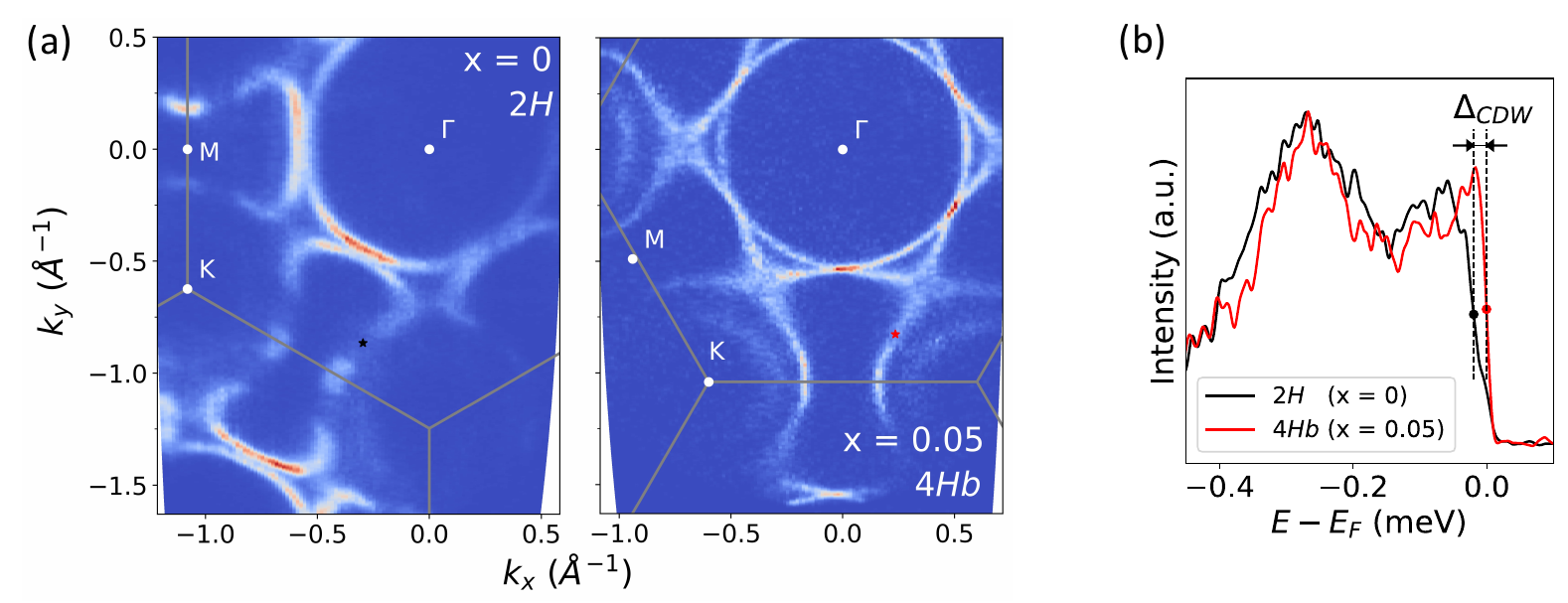}
    \caption{(a) Copy of the Fermi surfaces presented in Fig. 4(a) in the main text, with black and red stars indicating $k$-points at which EDCs were taken. 
    (b) Energy distribution curves for samples $x=0$ and 0.05, extracted at $k$-points indicated in panel (a) with black and red stars, respectively. The opening of the gap is assessed by a shift of the leading edge midpoint.}
    \label{fig:vts-sm-cdw}
\end{figure*}

\section{Suppression of CDW phase}

As further evidence of the gapped Fermi surface reconstruction in $2H$-\tas, in Fig.~\ref{fig:vts-sm-cdw}(b) we compare energy distribution curves (EDCs) between purely $2H$ ($x$ = 0) and $4Hb$ ($x = 0.05$) samples, taken at the gapped $k$-points of the Fermi surface. The curves show the opening of the CDW gap between $2H$ and $4Hb$ systems, providing additional evidence of suppression of the CDW phase in mixed layer structures. A residual Fermi step in the otherwise gapped $x=0$ spectrum resembles the CCDW phase of $2H$-\tase \cite{borisenko2008}. This could signal that the gap is centered slightly below $E_F$, or that some regions of the sample are ungapped due to disorder.

\section{Lindhard susceptibility}

\begin{figure*}[!htbp]% sm-lindhard
    \centering
    \includegraphics[width=0.95\linewidth]{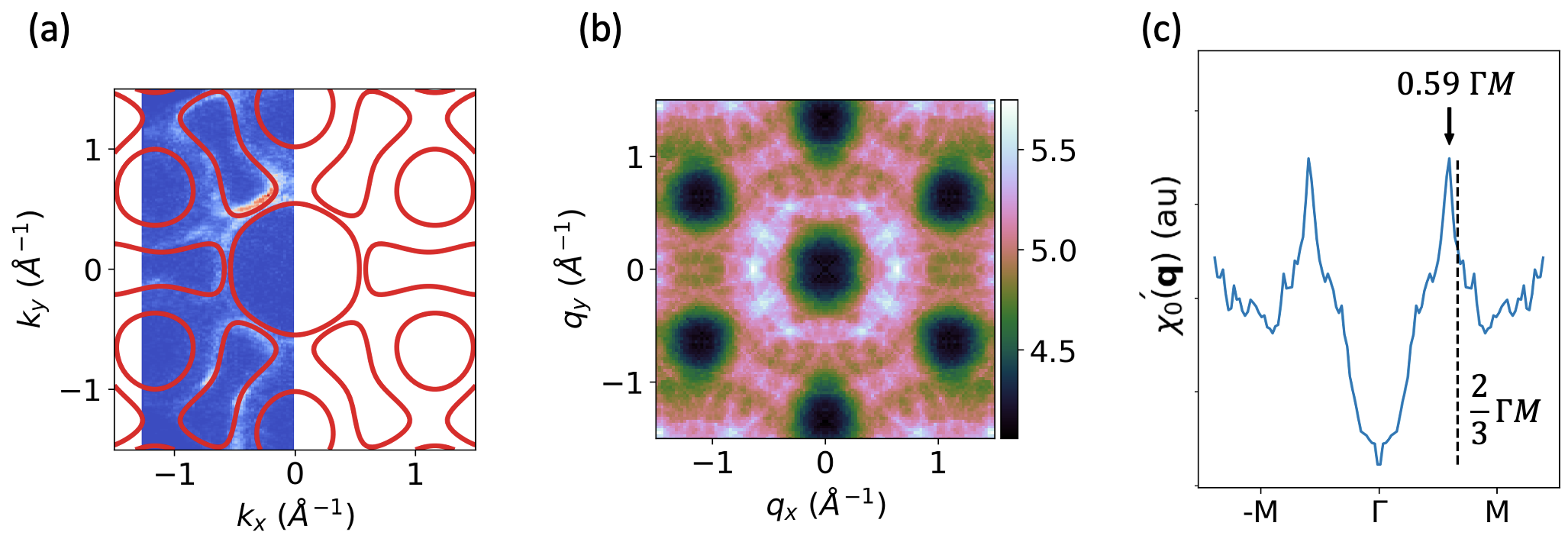}
    \caption{(a) Tight-binding model fitted to the $2H$-\tas data. 
    (b) Real part of the Lindhard susceptibility at $\omega \rightarrow 0$.
    (c) Slice of the Lindhard susceptibility $\chi_0'(q)$ along $\Gamma$-$M$ direction, indicating  peak in susceptibility.}
    \label{fig:vts-sm-lindhard}
\end{figure*}

To analyze the CDW in $2H$-\tas in more detail, we have used our ARPES data to compute the real part of the Lindhard susceptibility in the static limit ($\omega \rightarrow 0$). We follow the procedure applied by Inosov \textit{et al.} for $2H$ phases of \tase and NbSe$_2$ \cite{inosov2008}. We begin with finding tight-binding parameters that fit the acquired spectra, as presented in Fig.~\ref{fig:vts-sm-lindhard}(a). Using the obtained dispersions, we calculate the Lindhard function $\chi'_0(q)$ as described in Ref.~\cite{inosov2008}. Figure~\ref{fig:vts-sm-lindhard}(b) maps $\chi'_0$ as a function of momentum in the $q_x\text{-}q_y$ plane. Figure~\ref{fig:vts-sm-lindhard}(c) plots $\chi'_0(q)$ along the $\Gamma$-$M$ direction. In contrast to the expectations of a purely nesting-driven CDW model, the susceptibility is peaked at $0.59|\Gamma\mathrm{-}M|$, rather than the ordering wavevector $q_{\mathrm{CDW}}=\frac{2}{3}|\Gamma\mathrm{-}M|$. This result closely matches findings from $2H$-\tase and $2H$-NbSe$_2$ \cite{inosov2008}.

\section{Extracting electron-phonon coupling from ARPES data} 

We extract the electronic self-energy from the ARPES data following the general approach of Ref.~\cite{kordyuk2005}. For a given 2D ARPES  spectrum, we fit each momentum distribution curve at fixed energy $\omega=E-E_F$ with a Lorentzian to obtain the measured dispersion points $k(\omega)$ and spectral widths $\Gamma(\omega)$, defined in terms of full-width at half-maximum (FWHM). The real part of the self-energy relates to the difference between the linearized bare band and the renormalized dispersion, while the imaginary part is reflected in the spectral width. Namely we use the following formulas:

\begin{equation}
    \Sigma'(\omega)  = \omega - v_F^0 [k(\omega) - k_F]
    \label{eq:se_real}
\end{equation}
\begin{equation}
    \Sigma''(\omega)  = \frac{1}{2} \cdot v_F^0 \cdot \Gamma(\omega),
    \label{eq:se_imag}
\end{equation}

\noindent where $k_F = k(0)$ is the Fermi momentum obtained from the dispersion fitting, and $v_F^0$ is the bare band velocity. We use the fact that $\Sigma'$ and $\Sigma''$ are related to each other by the Kramers-Kronig (KK) transformation 

\begin{equation}
    \Sigma'(\omega) = \frac{1}{\pi} \int_{-\infty}^{+\infty} \frac{\Sigma''(\omega)}{\omega' - \omega} \mathrm{d}\omega',
    \label{eq:se_kk}
\end{equation}

\noindent in order to self-consistently obtain $v_F^0$ and thus the overall scaling of the real and imaginary components. Namely, we find the value of $v_F^0$ that minimizes the sum of the squared residuals between $\Sigma'(\omega)$ obtained from Eq.~(\ref{eq:se_real}) and from applying the KK tranform of Eq.~(\ref{eq:se_kk}).

To analyze the electron-phonon coupling based on the extracted self-energies, we adopt the procedure described in Ref.~\cite{wen2018}. Within the Migdal-Eliashberg theory \cite{carbotte2003}, the electron-phonon coupling constant is given by

\begin{equation}
    \lambda = 2\int_0^{\omega_{max}} \frac{\alpha^2F(\omega)}{\omega} \mathrm{d}\omega,
    \label{eq:lambda_se}
\end{equation}

where

\begin{equation}
    \alpha^2 F(\omega) = \frac{1}{\pi} 
    \frac{\mathrm{d}\Sigma''(\omega)}{\mathrm{d}\omega}.
\end{equation}

\begin{table}[ht!]
\caption{Values of electron-phonon coupling parameter (and propagated uncertainties) along the $\Gamma\text{--}K$, $\Gamma\text{--}M$ and $M\text{--}K$ high symmetry directions for samples $x=0$ and 0.05, calculated using two  methods: from mass enhancement ($\lambda_{v}$, see main text) and self-energy of the electron-phonon interactions ($\lambda_{se}$, described above).}
\label{tab:lambdas}
\begin{ruledtabular}
\begin{tabularx}{\linewidth}{cccc}
Sample                    & Direction     &$\lambda_{se}$  & $\lambda_{v}$ \\ \hline %& 
                          & $\Gamma\text{--}K$  & 0.494(14)      & 0.479(15) \\            %& 
                          & $\Gamma\text{--}M$  & 0.236(32)      & 0.211(28) \\            %& 
\multirow{-3}{*}{$x=0$}   & $M\text{--}K$       & 0.398(33)      & 0.342(49) \\ \hline     %& 
                          & $\Gamma\text{--}K$  & 0.511(30)      & 0.523(33) \\            %& 
                          & $\Gamma\text{--}M$  & 0.440(31)      & 0.422(21) \\            %& 
\multirow{-3}{*}{$x=0.5$} & $M\text{--}K$       & 0.400(25)      & 0.406(38)               %& 
\end{tabularx}
\end{ruledtabular}
\end{table}

Finally, the uncertainties associated with the above values are evaluated as

\begin{equation}\label{eq:delta_lambda}
    \Delta \lambda = 2 \sqrt {\sum_{\omega=0}^{\omega_{max}} \Bigg(\frac{\Delta\alpha^2F(\omega)}{\omega}\Bigg)^2} \cdot (5\mathrm{\ meV}),
\end{equation}

where

\begin{equation}
    \Delta \alpha^2F(\omega) = \frac{1}{\pi} \frac{\sqrt{[\Delta \Sigma''(\omega_{n-1})]^2+ [\Delta \Sigma''(\omega_{n+1})]^2}}{5 \mathrm{\ meV}}
    \label{eq:delta_a2f}
\end{equation}

and

\begin{equation}
    \Delta \Sigma'(\omega) = v_F^0 \cdot\Delta k(\omega) 
    \hspace{2cm}
    \Delta \Sigma''(\omega) = \frac{1}{2} \cdot v_F^0 \cdot \Delta \Gamma(\omega).
\end{equation}

\begin{figure*}[!htbp]% vts-lambdas-supplement
    \centering
    \includegraphics[width=0.3\linewidth]{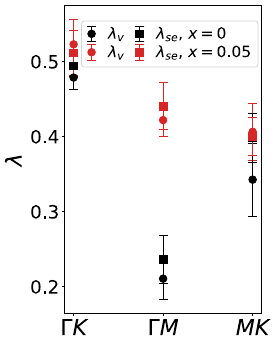}
    \caption{Electron-phonon coupling constant $\lambda$ along the $\Gamma\text{--}K$, $\Gamma\text{--}M$ and $M\text{--}K$ directions, for samples $x=0$ and 0.05. Values are obtained using two methods: comparing the change in the renormalized Fermi velocity ($\lambda_v$) and analyzing the self-energy of the electron-phonon interactions ($\lambda_{se}$).}
    \label{fig:vts-sm-lambdas}
\end{figure*}

\noindent The value of 5 meV in Eqs.~\ref{eq:delta_lambda} and \ref{eq:delta_a2f} is the energy step with which the analysis was carried out. $\Delta k(\omega)$ and $\Delta \Gamma(\omega)$ are standard deviations returned by the Lorentzian fitting procedure. The extracted $\lambda$ values together with their uncertainties are presented in Table~\ref{tab:lambdas} and Fig.~\ref{fig:vts-sm-lambdas}.

As a final check on the validity of the above methodology, we note that the electron-phonon coupling also relates directly to the mass enhancement of the coupled electrons via 

\begin{equation}\label{eq:lambda_v}
    \lambda=v_F^0/v_F-1,
\end{equation}

\noindent where $v_F$ is the Fermi velocity of the renormalized ``dressed'' electrons. We determine $v_F$ by linear fits to the $k(\omega)$ points in energy windows of 30, 20 and 40 meV below $E_F$ for cuts along $\Gamma\text{--}K$, $\Gamma\text{--}M$ and $M\text{--}K$, respectively, and obtain $v_F^0$ as described above. As shown in Table \ref{tab:lambdas} and Fig.~\ref{fig:vts-sm-lambdas}, the values of $\lambda$ computed from Eq.~\ref{eq:lambda_v} ($\lambda_{v}$) match the results of Eq.~\ref{eq:lambda_se} ($\lambda_{se}$) within the uncertainties. 

\section{Influence of $\lambda$ on superconducting $T_c$}

To illustrate the influence of the obtained electron-phonon coupling constants $\lambda$ on the superconducting $T_c$, we apply McMillan's formula

\begin{equation}
    T_c = \frac{\Theta}{1.45} \exp\left[-\frac{1.04(1+\lambda)}{\lambda - \mu^*(1+0.62\lambda)}\right],
   \label{eq:mcmillan}
\end{equation}

\noindent where $\Theta$ is a temperature scale corresponding to a phonon frequency cutoff, and $\mu^*$ is a Coulomb coupling term. We set $\Theta=580$~K based on a maximum calculated phonon energy of 50 meV \cite{hinsche2018}. For each sample, we average the electron-phonon coupling over the three measured momentum directions, giving $\lambda=$0.38 and 0.45 for $2H$ ($x=0$) and $4Hb$ ($x=0.05$), respectively. Based on these inputs, the experimental transition temperatures of 0.8 and 2.2 K for the $x=0$ and 0.05 samples are obtained for $\mu^*=$ 0.12 and 0.125, respectively. These $\mu^*$ values are very close to the value suggested in other works focusing on $2H$($1H$)-\tas \cite{lian2022, hinsche2018}. (It is reasonable to expect an increased value of $\mu^*$ in the $4Hb$ structure compared to the $2H$ phase, due to reduced screening. If $\mu^*$ is held constant, the calculated enhancement in $T_c$ due to the change in $\lambda$ alone is actually slightly larger than seen in the experiments.)

Naturally, this calculation rests on multiple assumptions and only crudely accounts for the variation of $\lambda$ within the Brillouin zone. Nevertheless, it illustrates how an overall strengthening of electron-phonon coupling on the order of $\sim20\%$ can go a long way in explaining the nearly threefold increase in $T_c$ in $4Hb$-\tas compared to the $2H$ phase.

\bibliography{bib}